\documentclass[aps,pre,reprint,groupedaddress,a4paper,onecolumn,notitlepage,showpacs]{revtex4-1}

\usepackage{lmodern,enumerate,natbib}
\usepackage{amsmath,amssymb,amsthm,amsfonts,mathrsfs}
\usepackage[colorlinks=true,linkcolor=blue,citecolor=blue,urlcolor=blue]{hyperref}
\usepackage{epsfig,epstopdf}

\newcommand{\natur}{\mathbb{N}}

\newcommand{\integer}{\mathbb{Z}}

\newcommand{\fp}[1]{\left\{#1\right\}}
\newcommand{\floor}[1]{\left\lfloor #1 \right\rfloor}
\renewcommand\Re{\operatorname{Re}}
\renewcommand\Im{\operatorname{Im}}

\newcommand{\comm}[2]{[#1, #2]}
\newcommand{\acomm}[2]{\{#1, #2\}}

\newcommand{\pauli}[1]{\sigma^{#1}}

\newcommand{\lgen}{\mathcal{L}}
\newcommand{\usup}{\mathcal{U}}

\begin{document}

\title{Markovian theory of dynamical decoupling by periodic control}

\author{Krzysztof Szczygielski}
\email{fizksz@ug.edu.pl}
\affiliation{Institute of Theoretical Physics and Astrophysics, University of Gdansk, Wita Stwosza 57, 80-952 Gdansk, Poland}

\author{Robert Alicki}
\email{fizra@ug.edu.pl}
\affiliation{Institute of Theoretical Physics and Astrophysics, University of Gdansk, Wita Stwosza 57, 80-952 Gdansk, Poland}

\date{\today}

\begin{abstract}
The recently developed formalism of Markovian master equations for quantum open systems with external periodic driving is applied to the theory of dynamical decoupling by periodic control. This new approach provides a more detailed quantitative picture of this phenomenon applicable both to traditional NMR spectroscopy and to recent attempts of decoherence reduction in quantum devices. For popular choices of bath spectral densities and a two-level system, exact formulas for the relaxation rates  are available. In the context of NMR imaging, the obtained formulas allow the extraction of a new parameter describing bath relaxation time from the experimental data.
\end{abstract}


\maketitle

\section{Introduction}

The historical origin of dynamical decoupling dates back to the beginning of 1950-ties when the ``spin-echo'' or ``refocusing'' phenomena were observed for nuclear magnetic resonance (NMR) \cite{Hahn:1950}. Namely, a properly designed sequence of pulses, which mimicks the time-reversal operation, reduced the decay of coherence for the NMR signal emitted by a large sample of nuclear spins. The ``static'' source of decoherence due to the inhomogeneity of the external magnetic field dominated over the ``dynamical'' decoherence caused by the interaction of a single spin with environmental noise. Therefore, the effective elimination of the former allowed the high-precision measurement of decoherence time $T_2$ due to the latter. Further progress in experimental techniques allowed the shortening of the repetition time of the pulses $T$ below the values of $T_2$, leading to the reduction of dynamical decoherence as well \cite{Haeberlen:1968}. In Ref. \cite{Haeberlen:1968} the elementary theory of this effect, called \emph{coherent averaging}, has been developed. 
\par
In the last two decades new advances in the fields of quantum information, quantum computing or, more generally, quantum control have led to the revival of this topic both on the theoretical and experimental sides (see Ref. \cite{ViolaLloyd:1998} and the book \cite{QEC}). Recent theoretical contributions concentrated mostly on the development of numerous, often quite sophisticated, decoupling schemes such as group-based designs, Eulerian design, or concatenated schemes which went far beyond simple "`bang-bang"' periodic decoupling. However, an approach based on coherent averaging applied to a generic bath could not give detailed quantitative predictions concerning the decay rates except for the universal condition for decoherence suppression by periodic control,
\begin{equation}
T \ll \frac{1}{\omega_{\mathrm{cut}}},
\label{nonM}
\end{equation}
where $\omega_{\mathrm{cut}}$ is supposed to be the cut-off frequency for the spectral density of the bath. Much more detailed information has been obtained for the exactly solvable case of Gaussian classical noise combined with $\pi$- rotations and for various decoupling schemes; see, e.g., Refs. \cite{CywinskiLutchyn:2008,BiercukBluhm:2011,KhodjastehSastrawan:2013}.
\par
The inequality \eqref{nonM} was often treated as a sign of the \emph{non-Markovian} character of the setting. Another picture invoked in the context of dynamical decoupling was the analogy to the \emph{quantum Zeno effect}--the freezing of dynamics due to frequent measurements. Perhaps the most insightful, from the physical point of view, is an explanation based on the Fermi golden rule. Here, a frequent periodic perturbation shifts the Bohr frequencies of the system into a frequency domain higher than that corresponding to relevant excitations in the bath \cite{Alicki:2006}. As a consequence, energy conservation suppresses efficient coupling to the bath.
\par
The aim of this paper is to apply the recently developed theory of quantum Markovian master equations for periodically driven systems to dynamical decoupling by \emph{periodic control operations}. This framework introduced in Ref. \cite{AlickiLidarZanardi:2006} and studied in Ref. \cite{Szczygielski:2014} was applied to a number of models of quantum engines and cooling schemes \cite{Gelbwaser:2013,Gelbwaser:2013a,Levy:2012,Kolar:2012,SzczygielskiGelbwaserAlicki:2013}. It is based on the combination of Davies' weak-coupling-limit technique \cite{Davies:1974} with Floquet's theory of periodic-in-time quantum Hamiltonians \cite{Howland:1979}. The main result of the formalism is the master equation written in the following form:
\begin{equation}\label{eq_MME2}
	\frac{d \rho_{\mathrm{int}}(t)}{dt} = \lgen \rho_{\mathrm{int}}(t),
\end{equation}
where $\rho_{\mathrm{int}}(t)$ is the time-dependent reduced density matrix of the system in the \emph{interaction picture} and $\lgen$ generates a completely positive dynamical semigroup, i.e., possesses the canonical Lindblad--Gorini--Kossakowski--Sudarshan structure \cite{GKS76,Lindblad76}. The latter property means that the dynamics is Markovian, which seems to contradict the popular opinion about the non-Markovian nature of the dynamical decoupling setting. 
\par
The origin of this controversy is again a very popular picture of the Markovian approximation invoking the short correlation time of the bath as a necessary condition. Actually, this correlation time can even be non-well-defined, as in the case of spontaneous emission where the electromagnetic bath correlations decay like $t^{-4}$ but, nevertheless, spontaneous emission is a perfect example of Markovian (exponential) decay. A more rigorous analysis shows that Markovian behavior is a consequence of the following conditions: (i) weak coupling to a large stationary bath with a quasicontinuous spectrum and approximately Gaussian correlations,
 (ii) a mild assumption like integrability of two-point correlation functions of the bath implying a continuous spectral density function, slowly varying on the scale of typical relaxation rate, and (iii) well-separated relevant Bohr frequencies of the system leading to an efficient time averaging of the Hamiltonian-induced oscillations during the typical relaxation time.
\par
As a consequence of the above conditions one should treat the equation \eqref{eq_MME2} as describing the \emph{coarse-grained-in-time} dynamics valid for times longer than the timescale determined by the Hamiltonian dynamics. In the case of fast periodic control this timescale is determined by the pulse repetition time. A rule of thumb for the applicability of the Markovian approximation is an essentially continuous energy spectrum of the bath and relaxation rates essentially lower than the relevant Bohr frequencies of the system, including pulse frequencies.
\par
The main advantage of the proposed formalism is the existence of explicit formulas for the effective decoherence and relaxation rates for periodically controlled systems depending on the details of the free-system Hamiltonian, the form of perturbation, the system-bath interaction, and the parameters of the bath. In simple but important cases even closed exact solutions can be obtained that allow a rigorous analysis of dynamical decoupling phenomena.

\section{Open system with periodic driving}

In this section we briefly present a theoretical framework of periodically driven open quantum systems, first in a general case (in Sect. \ref{subsection_OpenSystemsPeriodicGeneral}), then in a case of periodic kicking (in Sect. \ref{subsection_OpenSystemsPeriodicKicking}). We skip most of mathematical technicalities; the reader can find more details in Refs. \cite{AlickiLidarZanardi:2006,Szczygielski:2014,SzczygielskiGelbwaserAlicki:2013}. We also use the same units for energy and angular frequency, i.e., $\hbar\equiv 1$.
\par
We consider an open quantum-mechanical system S described, for simplicity, by a finite-dimensional Hilbert space and by a time-periodic Hamiltonian $H(t) = H(t +T)$. At this point we note that $H(t)$ must be understood as a properly renormalized, \emph{physical} Hamiltonian, already containing all Lamb-like shifts induced by interaction with the environment \cite{AlickiLendi:2006,HuelgaRivas:2012,Szczygielski:2014}. The system-bath interaction is parametrized as
\begin{equation}
	H_{\mathrm{int}} = \sum_{\alpha } S_{\alpha} \otimes R_{\alpha},
\label{Hint}
\end{equation}
where $S_{\alpha}$ and $R_{\alpha}$ are Hermitian operators of the system and bath, respectively. The initial state of the bath (with the average denoted by $\langle\cdots\rangle_B$) is invariant with respect to the free dynamics of the bath and satisfies $\langle R_{\alpha}\rangle_B =0$.
\par
Applying the standard derivation of the master equation in the interaction picture based on the weak-coupling-limit technique, one obtains equation \eqref{eq_MME2} with the generator
\begin{align}\label{eq_LindbladGenIP}
	\lgen\rho = &\sum_{\alpha , \beta}\sum_{\{\omega'\}} \gamma_{\alpha\beta}(\omega') \left( S_{\alpha}(\omega') \rho S_{\beta}(\omega')^{\dagger} - \frac{1}{2}\acomm{S_{\beta}(\omega')^{\dagger}S_{\alpha}(\omega')}{\rho} \right).
\end{align}
Here, operators $S_{\alpha}(\omega')$ are defined by the Fourier decomposition
\begin{equation}\label{eq_SoperatorExpansion}
	S_{\alpha}(t)\equiv U(t)^{\dagger}(t)S_{\alpha} U(t) = \sum_{\{\omega'\}}S_{\alpha}(\omega') e^{i\omega' t} \\
\end{equation}
where 
\begin{equation}\label{Hpropagator}
	\frac{d}{dt} U(t) = -i H(t) U(t), \quad U(0) = I,
\end{equation}
and a set of frequencies $\{\omega'\}$ will be determined later using Floquet theory. The relaxation and decoherence rates are determined by the Fourier transforms of bath correlations which, for any $\omega$, form a positively-defined matrix.
\begin{equation}\label{relaxation}
	\gamma_{\alpha\beta}(\omega) = \int\limits_{-\infty}^{\infty} e^{-i\omega t} \langle R_{\alpha}(t) R_{\beta}\rangle_B\, dt .
\end{equation}
Notice, that in both expressions \eqref{eq_SoperatorExpansion}, \eqref{relaxation} and in the similar ones later on, the time dependence of observables is always given in the Heisenberg picture, governed by internal Hamiltonian of the system or bath, respectively. The formula \eqref{relaxation} shows also why a short correlation time for the bath is not a necessary condition for the validity of the Markovian approximation. What is important is that the integral restricted to the interval $[- \tau , \tau]$ converges to the definite value, describing the relaxation matrix element. This convergence can be fast even for slowly decaying correlations if only the relevant Bohr frequencies $\{\omega\}$ are high enough.

\subsection{Application of Floquet theory}
\label{subsection_OpenSystemsPeriodicGeneral}

Applying Floquet theory to the unitary propagator $U(t)$, one obtains the following decomposition \cite{Szczygielski:2014}:
\begin{equation}\label{eq_PropagatorResolution}
	U(t) = P(t) e^{-i\bar{H}t},
\end{equation}
with $P(t)$ being periodic and unitary; $\bar{H}$ is a Hermitian \emph{averaged Hamiltonian}, satisfying the relation
\begin{equation}\label{eq_FloquetOperatorGeneral}
	U(T) = e^{-i\bar{H}T} .
\end{equation}
The \emph{Floquet operator} $U(T)$ is a unitary operator, which propagates a quantum-mechanical state $\psi(t)$ over the full cycle $[0, T]$. Both $U(T)$ and the averaged Hamiltonian $\bar{H}$ posses a common eigenbasis $\{\phi_{k}\}$; therefore, $U(T)\phi_{k} = e^{-\frac{i}{\hbar}\epsilon_{k}T}\phi_{k}$, where $\{\epsilon_{k}\}$ are called \emph{quasi-energies} of a system with a periodic Hamiltonian. This structure implies a particular form of Fourier decomposition \eqref{eq_SoperatorExpansion} given by
\begin{equation}\label{eq_SoperatorExpansion1}
	S_{\alpha}(t) = \sum_{q\in\integer} \sum_{\{\omega\}}S_{\alpha}(\omega,q) e^{i(\omega+q\Omega)t} \\
\end{equation}
where operators $S_{\alpha}(\omega,q)$ are subject to the relations 
\begin{align}\label{Srelations}
S_{\alpha}(\omega,q)^{\dagger} &= S_{\alpha}(-\omega,-q), \nonumber \\ \comm{\bar{H}}{S_{\alpha}(\omega,q)} &= \omega S_{\alpha}(\omega,q).
\end{align}
The set of frequencies $\{\omega'\}$ in Eq. \eqref{eq_SoperatorExpansion} is labeled by two elements $\{\omega'\} \equiv \{\omega + q\Omega ; \, q = 0,\,\pm 1 ,\, \pm 2,\,\dots\}$
where $\{\omega\} = \{\epsilon_{k}-\epsilon_{l} \}$ are Bohr-Floquet quasifrequencies, and $\Omega = 2\pi/T$ . The harmonics $ q\Omega$ correspond to the energy quanta $q\Omega$ which are exchanged with the external source of periodic driving.
\subsection{Properties of Master equations for periodic driving}
By using the decomposition \eqref{eq_SoperatorExpansion} one can rewrite the interaction-picture generator \eqref{eq_LindbladGenIP} in a the form
\begin{align}\label{eq_LindbladGenIP1}
	\lgen\rho = &\sum_{\alpha , \beta}\sum_{q\in\integer}\sum_{\{\omega\}} \gamma_{\alpha\beta}(\omega+q\Omega) \left( S_{\alpha}(\omega,q) \rho S_{\beta}(\omega,q)^{\dagger} - \frac{1}{2}\acomm{S_{\beta}(\omega,q)^{\dagger}S_{\alpha}(\omega,q)}{\rho} \right).
\end{align}
From the properties \eqref{Srelations} it follows that the generator \eqref{eq_LindbladGenIP} independently transforms diagonal and off-diagonal elements of $\rho$, computed in the eigenbasis of the averaged Hamiltonian $\bar{H}$.
\par
It is often useful to employ the Markovian master equation in the \emph{Schr\"{o}dinger picture} which possesses the following structure:
\begin{equation}\label{eq_MME_Schroedinger}
	\frac{d\rho(t)}{dt} = -i \comm{H(t)}{\rho(t)} + \left(\usup(t) \lgen \, \, \usup(t)^{\dagger}\right) \rho(t)
\end{equation}
with the unitary superoperator $\usup(t) \rho = U(t) \rho \, U(t)^{\dagger}$. The properties \eqref{Srelations} imply also a very useful factorization property of the dynamics governed by \eqref{eq_MME_Schroedinger}
\begin{equation}\label{solution}
	\rho(t) = \usup(t) e^{t\lgen} \rho(0),
\end{equation}
which allows to discuss separately the decoherence or dissipation effects described by $\lgen$ and the unitary evolution $\usup(t)$.
\subsection{Periodic kicking}
\label{subsection_OpenSystemsPeriodicKicking}

Now we focus on a very special case of periodic Hamiltonian; namely, the so-called \emph{kicked Hamiltonian}, which can be formally given as
\begin{equation}\label{eq_PeriodicallyKickedHamiltonian}
	H(t) = H_0 + \lambda W \sum_{k=-\infty}^{\infty} \delta(t-kT),
\end{equation}
where $\lambda$ is a cumulated kicking strength and $H_0$ and $W$ are Hermitian operators. As was already suggested, the key point for open-system analysis lays in the Floquet operator $U(T)$ and its spectrum, which allows one to write the appropriate formula for the interaction-picture generator $\lgen$ \eqref{eq_LindbladGenIP}. A formal expression for the Floquet operator in the case of periodically kicked systems has been known for quite a long time already (see, e.g., Refs. \cite{Combescure:1990,McCawEtAl:2005,McCawMcKellar:2005,McCawPhDthesis:2005} and references therein). Because of the singular nature of $H(t)$, heuristically speaking, one is interested in a unitary evolution taking the state $\psi(t)$ not from time 0 to $T$, but rather from a time $0^{+}$ \emph{just after} the time of the first kick to a time $T^{+}$ \emph{just after} the time of the second kick. In the infinitesimally short interval $[T^{-},T^{+}]$, the system experiences a kick of infinite strength by the operator $\lambda W$, so, effectively, the influence from $H_0$ can be omitted. As a consequence the Floquet operator is given by
\begin{equation}\label{eq_KickedFloquetOperator}
	U(T) = e^{-i\lambda W} e^{-i H_{0}T} \equiv e^{-i \bar{H}T} .
\end{equation}
The time-dependent unitary $U(t)$ can be expressed in terms of two functions
 $\fp{x}$ and $\floor{x}$, called the \emph{fractional part} and the \emph{floor function}, respectively (see the Appendix), i.e.,
\begin{subequations}
\begin{align}
	&\floor{x} = \max{\{k \in \integer : k \leqslant x\}}, \\
	&\fp{x} = x - \floor{x}.
\end{align}
\end{subequations}
The $\fp{x}$ function is also known as the sawtooth function and is periodic with period 1, which directly implies that function $\fp{\frac{t}{T}}$ will be of period $T$. But this allows one to write,
\begin{equation}\label{eq_PropagatorSawtooth}
	U(t) = e^{-i H_0 T \fp{\frac{t}{T}}} e^{i\bar{H} T \fp{\frac{t}{T}}} e^{-i \bar{H}t}
\end{equation}
where $\bar{H}$ is given by \eqref{eq_KickedFloquetOperator}. This is naturally consistent with the representation \eqref{eq_PropagatorResolution} implied by the Floquet theorem, such that $P(t) = e^{-i H_0 T \fp{\frac{t}{T}}} e^{i\bar{H} T \fp{\frac{t}{T}}}$ is a periodic time-dependent operator; because the sawtooth function is periodic, this guarantees that results obtained in Ref. \cite{Szczygielski:2014} apply and the whole construction presented in Sec. \eqref{subsection_OpenSystemsPeriodicGeneral} can be employed. It is important that, from the very construction, function $U(t)$ given as Eq. \eqref{eq_PropagatorSawtooth} is discontinues at times $t = nT$, which is a direct consequence of the singular form of the Hamiltonian. Therefore, the resulting completely positive dynamics $\Lambda(t) = \usup(t) e^{t\lgen}$ will be discontinuous as well.
\section{Examples of "bang-bang" control}
In order to illustrate the power of the above formalism we discuss the simplest, but practically very useful, model of a two-level system (TLS) which can be physically realized as, e.g., spin $1/2$, superconducting qubit, quantum dot or fluorescence center in solid. We restrict ourselves to the most popular case of periodic control by short pulses which can be treated within the formalism of kicked dynamics. The chosen details of the Hamiltonian and the interpretation of the model parameters
are motivated by the NMR applications but obviously can be extended to other examples.
\par
We consider a two-level system described by the standard Pauli matrices.
The system unperturbed Hamiltonian is given by $\frac{1}{2} \omega_{0} \sigma^3$ and the control is executed by the external field oscillating with the angular frequency $\omega_{\mathrm{ext}}$ and with the envelope $f(t)$. The total TLS Hamiltonian in the rotating wave approximation reads
\begin{equation}
H'(t)= \frac{1}{2} \omega_{0} \pauli{3} + f(t)\bigl( e^{-i\omega_{\mathrm{ext}} t}\pauli{+} + e^{i\omega_{\mathrm{ext}} t}\pauli{-}\bigr)  .
\label{Hint2}
\end{equation}
The generic interaction with the environment can be written as
\begin{equation}
H_{\mathrm{int}}	= \vec{\sigma} \cdot \vec{R}
	\label{generic_int}
\end{equation}
where $\vec{R} \equiv (R_1, R_2 , R_3)$ represents three observables of the environment which in the NMR context, are usually chosen to be statistically independent. Then each component of the interaction can be treated independently and the cross correlations between different $R_j$ can be neglected.

\subsection{Example 1: Two-level system with ``longitudinal" coupling}

First, we consider the system-bath interaction Hamiltonian \eqref{Hint} of the form
\begin{equation}
H_{\mathrm{int}}= \pauli{3}\otimes R_3 .
\label{Hint1}
\end{equation}
This type of coupling, called in the NMR context the \emph{spin--spin interaction}, often dominates and hence, is interesting to be studied separately from the others.\\

In the absence of external control, i.e., for $f(t) = 0$, this model describes the \emph{pure decoherence} of the TLS with the decoherence time traditionally denoted by $T_2$ and given in the Markovian approximation by
\begin{equation}
T_{2} = \frac{2}{\gamma_{\parallel}(0)}, \quad \gamma_{\parallel}(\omega) = \int\limits_{-\infty}^{\infty} e^{-i\omega t} \langle R_3(t) R_3\rangle_B\, dt .
\label{T2}
\end{equation}
\vskip\baselineskip
In order to simplify the analysis we describe the system in the ``rotating frame," i.e., we perform a unitary transformation $\mathcal{U}_{0}^{\dagger}(t)$ defined as
\begin{equation}
	\mathcal{U}_{0}^{\dagger}(t)A = e^{\frac{1}{2}it\omega_{\mathrm{ext}}\pauli{3}} A e^{-\frac{1}{2}it\omega_{\mathrm{ext}}\pauli{3}}
\end{equation}
for an operator $A$, which will effectively switch us to the Heisenberg picture with respect to $\frac{1}{2}\omega_{\mathrm{ext}}\pauli{3}$. In this rotating frame the dynamics will be described by transformed Hamiltonian 
\begin{equation}\label{transformed_H}
H^{\prime\prime}(t) = \frac{\Delta}{2} \pauli{3} + f(t)\pauli{1}, 
\end{equation}
where 
\begin{equation}\label{Delta}
\Delta = \omega_{0}-\omega_{\mathrm{ext}}
\end{equation}
stands for the detuning parameter. 
\par
Now we assume that the envelope $f(t)$ is an infinite sequence of short pulses of duration much shorter than their separation time $T$ which is, on the other hand, much shorter than the decoherence time $T_2$. Therefore, we can replace pulses by Dirac deltas to obtain the final effective TLS Hamiltonian in the rotating frame
\begin{equation}\label{eq_PeriodicallyKickedHamiltonian1}
	H(t) = \frac{\Delta}{2}\pauli{3} + \lambda \left[\sum_{k=-\infty}^{\infty} \delta(t-kT)\right]\pauli{1}.
\end{equation}
We choose $\lambda = \pi/2$ (``magic angle," $\pi$-rotation) what drastically simplifies the description, but the exact solutions, albeit very involved, are also available for any $\lambda$. The Floquet operator $U(T)$ is, according to Eq. \eqref{eq_KickedFloquetOperator}, given as
\begin{align}
	U(T) &= e^{-\frac{i\pi}{2}\pauli{1}} e^{-\frac{iT\Delta}{2}\pauli{3}} \\
	&= -i \left( \begin{array}{cc} 0 & e^{\frac{1}{2}i\Delta T} \\ e^{-\frac{1}{2}i\Delta T} & 0 \end{array}\right) = e^{-iT \bar{H}}
\end{align}
and the Floquet basis is
\begin{equation}
	\phi_{1} = \frac{1}{\sqrt{2}} \left(\begin{array}{c} e^{\frac{1}{2}i\Delta T} \\ 1 \end{array}\right), \quad \phi_{2} = \frac{1}{\sqrt{2}} \left(\begin{array}{c} -e^{\frac{1}{2}i\Delta T} \\ 1 \end{array}\right).
\end{equation}
Quasienergies, i.e., eigenvalues of $\bar{H}$, are found to be $\epsilon_{1,2} = \pm \pi/(2T)$ and quasifrequencies are simply $\omega_{\pm} = \pm \pi/T $. The unitary propagator $U(t)$ can be easily found by direct application of Eq. \eqref{eq:propagator2}; in the Floquet basis $\{\phi_{1,2}\}$ it is given by
\begin{equation}
	U(t) = \left( \begin{array}{cc} e^{\frac{1}{2}i\pi\floor{\frac{t}{T}}} \cos{\left(\frac{\Delta T}{2}\fp{\frac{t}{T}}\right)} & i e^{-\frac{1}{2}i\pi\floor{\frac{t}{T}}} \sin{\left(\frac{\Delta T}{2}\fp{\frac{t}{T}}\right)} \\ i e^{\frac{1}{2}i\pi\floor{\frac{t}{T}}} \sin{\left(\frac{\Delta T}{2}\fp{\frac{t}{T}}\right)} & e^{-\frac{1}{2}i\pi\floor{\frac{t}{T}}} \cos{\left(\frac{\Delta T}{2}\fp{\frac{t}{T}}\right)} \end{array} \right) .
\end{equation}
This allows to obtain the Heisenberg picture evolution of $S = \pauli{3}$,
\begin{align}
	\pauli{3}(t) = \sum_{q=-\infty}^{\infty} &\left( S(+;q) e^{i\left(\frac{\pi}{T}+q\Omega\right)t} + S^{\dagger}(-;q) e^{-it\left(\frac{\pi}{T}+q\Omega\right)t} \right)
\end{align}
with matrices
\begin{subequations}
\begin{align}
	S(+;q) &= \frac{2i}{\pi(2q+1)} \left( \begin{array}{cc} 0 & 0 \\ 1 & 0 \end{array} \right), \\
	S(-;q) &= -\frac{2i}{\pi(2q-1)} \left( \begin{array}{cc} 0 & 1 \\ 0 & 0 \end{array} \right).
\end{align}	
\end{subequations}
We can cast Markovian master equation \eqref{eq_MME2} into an infinite sum
\begin{equation}\label{eq_LindbladianSeries}
	\mathcal{L}\rho = \sum_{q=-\infty}^{\infty} \mathcal{L}_{q} \rho
\end{equation}
where $\mathcal{L}_{q}$ is defined as
\begin{equation}
	\mathcal{L}_{q}\rho = \frac{2}{\pi^2}\left( \begin{array}{cc} -\frac{2\gamma\left(\left(\frac{1}{2}+q\right)\Omega\right)}{(1+2q)^2}\rho_{11} + \frac{2\gamma\left(\left(-\frac{1}{2}+q\right)\Omega\right)}{(1-2q)^2} \rho_{22} & -\left(\frac{\gamma\left(-\left(\frac{1}{2}+q\right)\Omega\right)}{(1-2q)^2} + \frac{\gamma\left(\left(\frac{1}{2}+q\right)\Omega\right)}{(1+2q)^2}\right) \rho_{12} \\
	- \left( \frac{\gamma\left(\left(-\frac{1}{2}+q\right)\Omega\right)}{(1-2q)^2} + \frac{\gamma\left(\left(\frac{1}{2}+q\right)\Omega\right)}{(1+2q)^2} \right)\rho_{21} & \frac{2\gamma\left(\left(\frac{1}{2}+q\right)\Omega\right)}{(1+2q)^2}\rho_{11} - \frac{2\gamma\left(\left(-\frac{1}{2}+q\right)\Omega\right)}{(1-2q)^2} \rho_{22}.
	\end{array} \right)
\end{equation}
The form \eqref{eq_LindbladianSeries} raises a natural question regarding summability of the infinite series; however it is easy to show that since $\gamma(\omega) \geqslant 0$ the series indeed converges whenever $\gamma (\omega)$ is bounded.
\par
We consider now the Lorentzian spectral density corresponding to the exponential decay of bath correlations with the decay time $\tau_{\mathrm{c}}$
\begin{equation}
	\gamma_{\parallel}(\omega) = \frac{2}{T_2}\frac{1}{1+\tau_{\mathrm{c}}^{2} \omega^{2}},
\label{lorentz}	
\end{equation}
where $T_2$ is the standard pure decoherence or dephasing time measured in the absence of modulation. The formula \eqref{lorentz}	 is a reasonable high-temperature approximation for the ``spin-spin" interaction with the bath, often used for NMR. 
\par
\begin{figure}[htbp]
	\centering
		\includegraphics{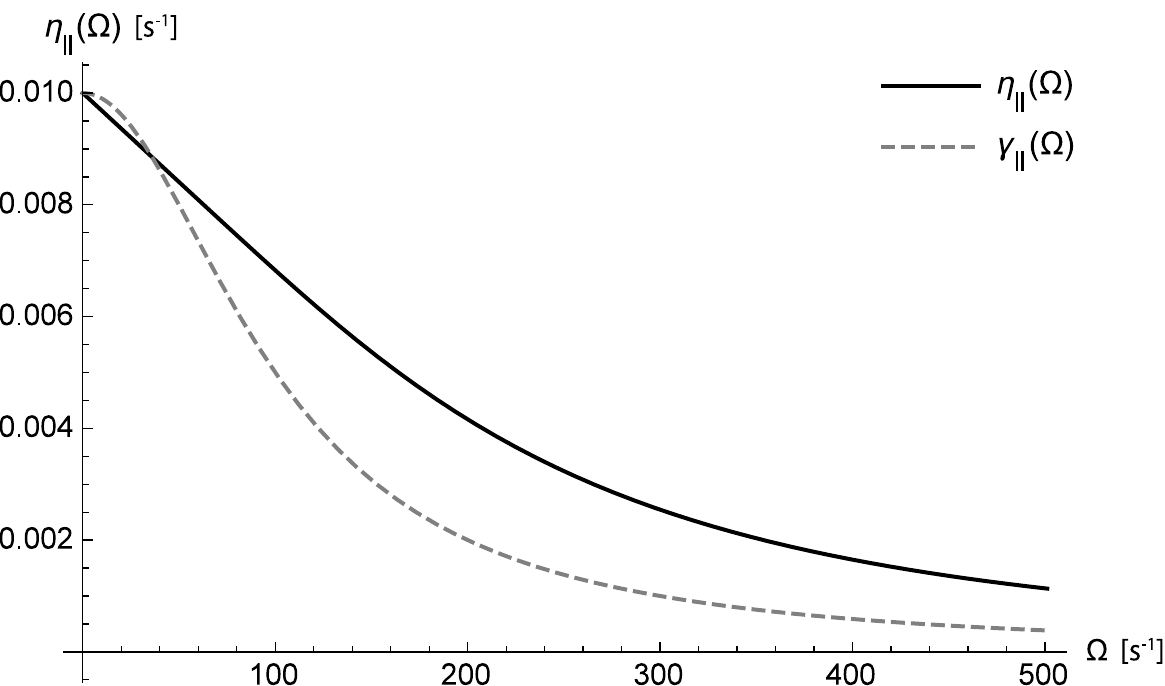}
	\caption{Comparison between the relaxation rate $\eta_{\parallel}(\Omega)$ (black) and spectral density $\gamma_{\parallel}(\Omega)$ (gray, dashed) for the longitudinal coupling.}
	\label{fig:GammaT2}
\end{figure}

With the choice \eqref{lorentz}	the summation in the formula \eqref{eq_LindbladianSeries} can be executed and produces the interaction picture generator
\begin{equation}\label{Lindblad_matrix}
	\lgen\rho = -\eta_{\parallel}(T)\left( \begin{array}{cc} \rho_{11} - \rho_{22} & \rho_{12} \\ \rho_{21} & \rho_{22} - \rho_{11} \end{array}\right),
\end{equation}
where the relaxation rate $\eta_{\parallel}(T)$ is expressed by the formula
\begin{equation}
	\eta_{\parallel}(T) = \frac{1}{T_2}\left( 1-\frac{2\tau_{\mathrm{c}}}{T} \tanh{\frac{T}{2\tau_{\mathrm{c}}}}\right).
\label{decoh_rate}	
\end{equation}
Figure \ref{fig:GammaT2} shows a plot of the relaxation rate $\eta_{\parallel}$ as a function of kicking frequency $\Omega$ compared to the Lorentzian spectral density \eqref{lorentz}.
\par
Now one can solve the interaction picture Markovian master equation and then transform the solution $\rho_{\mathrm{int}}(t)= e^{t\mathcal{L}}\rho(0) $ back to the Schr\"{o}dinger picture in the laboratory frame to obtain the exact expression for system's density operator in the original eigenbasis of $\pauli{3}$ operator,
\begin{equation}
	\rho(t) = \mathcal{U}_{0}(t) \left( \mathcal{U}(t)e^{t\mathcal{L}}\rho(0) \right)
\end{equation}
with explicit matrix elements $\rho_{ij}(t)$, $i,j\in\{1,2\}$ given by formulas ($\eta\equiv\eta_{\parallel}(T)$)
\begin{subequations}\label{eq:DensityOperatorSchroedinger}
	\begin{align}
		\rho_{11}(t) &= \frac{1}{2} \left[ 1+(-1)^{\floor{\frac{t}{T}}}e^{-\eta t} (\rho_{11}(0)-\rho_{22}(0)) \right], \\
				\rho_{21}(t) &= \frac{1}{2} e^{-2\eta t} e^{-i \left(\Delta T \floor{\frac{t}{T}}-\omega_{0}t\right)} \left( e^{-i\Delta T}\rho_{12}(0)+\rho_{21}(0) \right) \\
		&- \frac{1}{2}(-1)^{\floor{\frac{t}{T}}} e^{-\eta t} e^{-i\left(\Delta T \left( 1+\floor{\frac{t}{T}} \right)-\omega_{0}t\right)} (\rho_{12}(0)-e^{i\Delta T}\rho_{21}(0)) \nonumber\\
		\rho_{22}(t) &= 1 - \rho_{11}(t), \qquad \rho_{12}(t) = \overline{\rho_{21}(t)}.
	\end{align}
\end{subequations}
\vskip\baselineskip
We rewrite the solution \eqref{eq:DensityOperatorSchroedinger} in terms of Bloch ball parametrization for the TLS density matrix
\begin{equation}
	\rho = \frac{1}{2} \left[ I + \vec{x} \cdot \vec{\sigma}\right]
	\label{Bloch_ball}
\end{equation}
with components $x_{i}(t)$ of the Bloch vector given explicitly as
\begin{subequations}
	\begin{align}
		&x_{1} = \rho_{12} + \rho_{21} = 2\Re\rho_{21}, \\
		&x_{2} = i\left( \rho_{12} - \rho_{21} \right) = 2\Im\rho_{21}, \\
		&x_{3} = \rho_{11} - \rho_{22} = 2\rho_{11} - 1.
	\end{align}
\end{subequations}
By using the definition 
\begin{equation}
\phi(t) = \omega_{\mathrm{ext}} t + \Delta T\fp{\frac{t}{T}}_c ,
	\label{phi_phase}
\end{equation}
where $\fp{x}_c \equiv \fp{x} - 1/2$ can be called the \emph{centered sawtooth function}, one obtains the solution of Eq. \eqref{eq:DensityOperatorSchroedinger} in the form
\begin{subequations}
	\begin{align}
		x_{1}(t) &= e^{-2\eta t} \cos{\phi(t)}\left( x_{1}(0) \cos{\phi(0)}+x_{2}(0) \sin{\phi(0)} \right) \\
		&+ (-1)^{\floor{\frac{t}{T}}}e^{-\eta t} \sin{\phi(t)}( x_{1}(0) \sin{\phi(0)} - x_{2}(0) \cos{\phi(0)} ) \nonumber \\
		x_{2}(t) &= e^{-2\eta t} \sin{\phi(t)}\left( x_{2}(0) \sin{\phi(0)} + x_{1}(0) \cos{\phi(0)} \right) \\
		&+ (-1)^{\floor{\frac{t}{T}}}e^{-\eta t} \cos{\phi(t)}(x_{2}(0) \cos{\phi(0)} - x_{1}(0) \sin{\phi(0)}) \nonumber \\
		x_{3}(t) &= (-1)^{\floor{\frac{t}{T}}} e^{-\eta t} x_{3}(0).
	\end{align}
	\label{bloch_evolution}
\end{subequations}
One can decompose the evolution of Bloch vector into three types of motion: (a) rotation with the angular frequency $\omega_{\mathrm{ext}}$ perturbed by the  phase modulation of a sawtooth shape, (b) exponential decay with two decay rates, $\eta_{\parallel}(T)$ and $2\eta_{\parallel}(T)$, and (c) inversion of the slowly decaying component performed at the times $nT$. The exponential decay of all components of Bloch vector drives the TLS to the final maximally mixed state.
\subsubsection{Spin-echo phenomenon}
In real experiments one measures the NMR signal produced by the $x_1(t), x_2(t)$ components averaged over a large sample of individual nuclear spins. Because the external magnetic field which defines the Larmor frequency $\omega_0$ is not perfectly homogeneous the formula for phase \eqref{phi_phase} contains a deterministic component $\omega_{\mathrm{ext}} t$ and the essentially random detuning $\Delta = \omega_0 - \omega_{\mathrm{ext}}$. Averaging the terms, which describe rotation in the $x_1 x_2$ - plane, with respect to fluctuations of $\Delta $ one obtains
\begin{equation}
\langle \cos{\phi(t)}\rangle = \left\langle \cos{\left(\Delta T\fp{\frac{t}{T}}_c\right)}\right\rangle \cos{\omega_{\mathrm{ext}}t} -\left\langle \sin{\left(\Delta T\fp{\frac{t}{T}}_c\right)}\right\rangle \sin{\omega_{\mathrm{ext}}t}
\label{detunin_av}	
\end{equation}
\begin{equation}
\langle \sin{\phi(t)}\rangle = \left\langle \cos{\left(\Delta T\fp{\frac{t}{T}}_c\right)}\right\rangle \sin{\omega_{\mathrm{ext}}t} +\left\langle \sin{\left(\Delta T\fp{\frac{t}{T}}_c\right)}\right\rangle \cos{\omega_{\mathrm{ext}}t} .
\label{detunin_avsin}	
\end{equation}
For times $t= (n +1/2) T $ the function $\fp{\frac{t}{T}}_c$ is equal to zero, which implies that
\begin{equation}
\left\langle \cos{\phi\left(\left(n +\frac{1}{2}\right) T\right)}\right\rangle = \cos{\left(\omega_{\mathrm{ext}}\left(n +\frac{1}{2}\right) T\right)} ,
\label{detunin_av1}	
\end{equation}
and similarly
\begin{equation}
\left\langle \sin{\phi\left(\left(n +\frac{1}{2}\right) T\right)}\right\rangle = \sin{\left(\omega_{\mathrm{ext}}\left(n +\frac{1}{2}\right) T\right)}.
\label{detunin_av2}	
\end{equation}
Hence, the effect of magnetic field inhomogeneity is undone in those particular moments of time. This is the spin-echo phenomenon and the formulas \eqref{phi_phase}--\eqref{detunin_avsin}	allow us to compute the shape of the NMR signal for a given model of magnetic-field variations.
\subsubsection{Freezing of dynamics for fast pulses}
\begin{figure}[htbp]
	\centering
		\includegraphics[width=1.00\columnwidth]{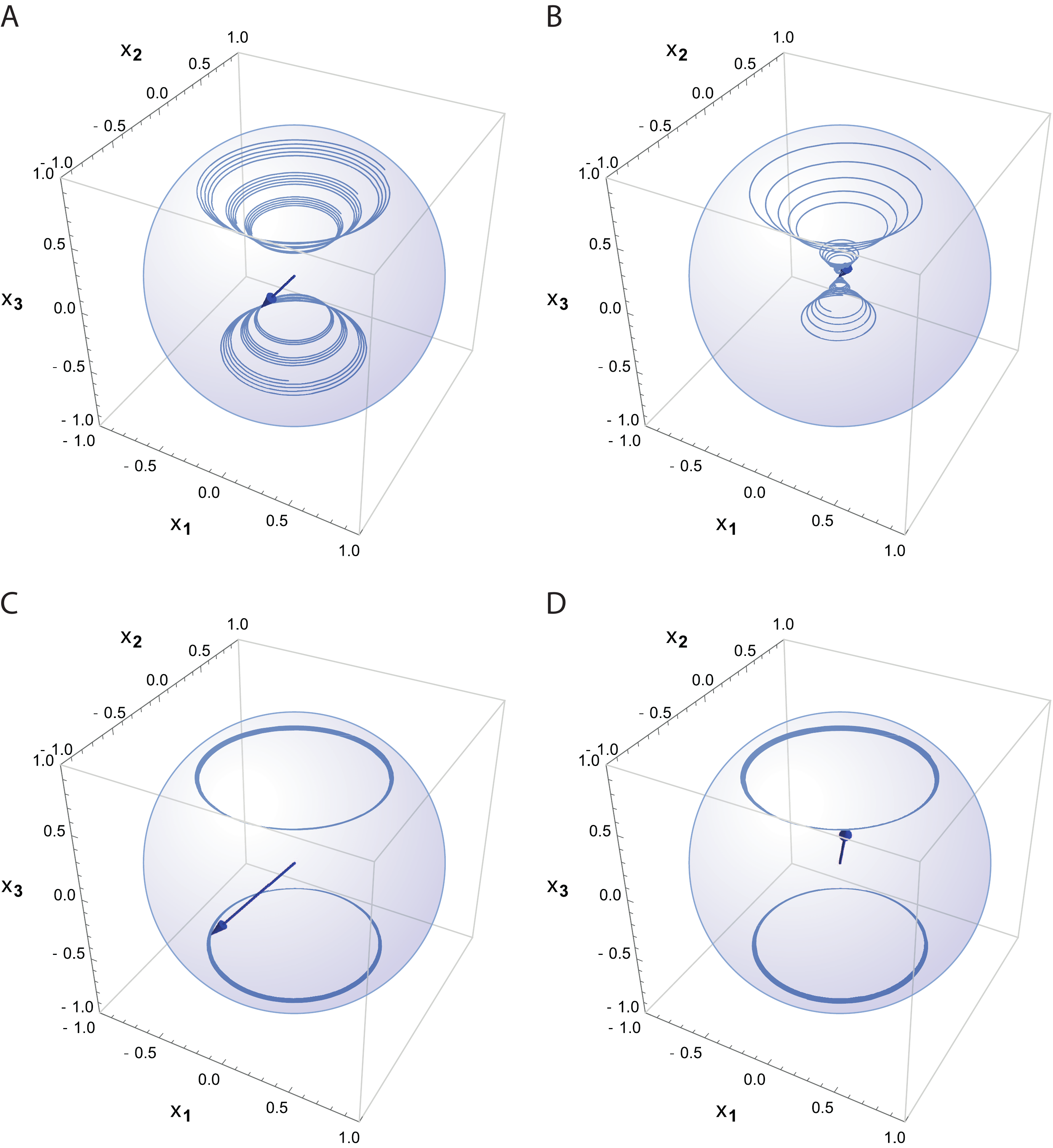}
	\caption{(Color online) Trajectories of TLS's Bloch vector $\vec{x}(t)$ for longitudinal coupling and Lorentzian bath spectrum. (a) Example of evolution with parameters $T_2 = 6.5\times 10^{-3}$, $\tau_{\mathrm{c}}=18.7$, $\omega_{0}=18.8$, and $T=1.5$. (b) $T_2$ is reduced by a factor of five ($T_2 = 1.3\times 10^{-3}$), producing a more unstable evolution and a faster approach to the stationary maximally mixed state. (c) The bath correlation time $\tau_{\mathrm{c}}$ is multiplied by a factor of five ($\tau_{\mathrm{c}} = 120$) keeping $T_2$ constant. (d) The period of modulation $T$ is decreased five times ($T = 0.3$)--both panels (c) and (d) result in much more stable evolution with long decoherence time. All plots were generated under $\Delta = 0$ condition.}
	\label{fig:Trajectories}
\end{figure}

Due to the spin-echo phenomenon, in NMR experiments, for the pulse period $T \gg \tau_c$, the measured decoherence time is equal to $T_{2}^{\prime}(T)= \eta_{\parallel}(T)^{-1} \simeq T_2$, where $T_2$ is the ideal decoherence time for a single spin in the absence of pulses [see formula \eqref{decoh_rate}]. Increasing the frequency of pulses to the values higher than $\tau_c^{-1}$ ($T \ll \tau_c$), one observes suppression of the relaxation rate $\eta_{\parallel}(T)$ (see Fig. \ref{fig:GammaT2}), which causes ``freezing" of dynamics--a phenomenon called \emph{decoupling by ``bang-bang'' control}. Figure \ref{fig:Trajectories} illustrates this effect showing the evolution of Bloch vector \eqref{bloch_evolution} for different parameters of the model.

\subsection{Example 2: Two-level system with ``transverse" coupling}

Here we examine a more involved case of directions transverse to the constant external magnetic field which appear in the NMR spin-bath coupling. The system's Hamiltonian is the same as in the previous case \eqref{eq_PeriodicallyKickedHamiltonian1}, and the ``magic angle'' condition $\lambda = \pi/2$ still applies. The Floquet operator $U(T)$ and the evolution operator $U(t)$ are also the same. The interaction Hamiltonian $H_{\mathrm{int}}$ is now, however, of a form
\begin{equation}
	H_{\mathrm{int}} = \pauli{1}\otimes R_{1} + \pauli{2} \otimes R_{2},
	\label{coupling_perpendicular}
\end{equation}
and in the context of NMR describes the ``spin-lattice" coupling.
\subsubsection{The case of no control, \texorpdfstring{$f(t) = 0$}{f(t) = 0}.}
In the absence of external control, the two terms in \eqref{coupling_perpendicular} lead to similar effects and act additively. For a thermal bath at the inverse temperature $\beta$, TLS thermalizes with two decay times: $T_1$--for the diagonal elements (in $\pauli{3}$ basis) and $2T_1$--for the off-diagonal elements of the density matrix.
$T_1$ is determined by the joint spectral density of a bath, satisfying the Kubo-Martin-Schwinger condition
\begin{equation}
 \gamma_{\perp}(\omega) = \sum_{j=1}^2\int\limits_{-\infty}^{\infty} e^{-i\omega t} \langle R_j(t) R_j\rangle_B\, dt , \quad \gamma_{\perp}(-\omega) = e^{-\beta\omega} \gamma_{\perp}(\omega),
\label{gamma_perp}
\end{equation}
as
\begin{equation}
T_{1} = \bigl[ (1 + e^{-\beta\omega_0})\gamma_{\perp}(\omega_0)\bigr]^{-1}.
\label{T1}
\end{equation}
The standard model for the spectral density of the acoustic phonon bath is given by the regularized expression 
\begin{equation}
\gamma_{\perp}(\omega) = \frac{A\omega^{3}}{1-e^{-\beta \omega}} e^{-\frac{\omega}{\omega_{\mathrm{cut}}}}
\label{phonon_bath}
\end{equation}
where $A$ is an effective coupling strength and $\omega_{\mathrm{cut}}$ is a cutoff corresponding to the Debye frequency. The similar formula can be used for the case of electromagnetic bath with dipole coupling which can be relevant for TLS models describing quantum dots, superconducting qubits, etc.
\par
If a bath with a spectral density of the type \eqref{phonon_bath} is also coupled by the $\pauli{3}$ term; it does not contribute to the decoherence time, because $\gamma_{\perp}(0) = 0$. Therefore, under the joint influence of both types of baths the ``spin-lattice'' relaxation time is given by Eq. \eqref{T1} and the decoherence time $T_{2}^{\prime}$ is finally given by
\begin{equation}
\frac{1}{T^{\prime}_{2}} = \frac{1}{T_2} + \frac{1}{2T_1} = \frac{1}{2}\Bigl(\gamma_{\parallel}(0) + (1 + e^{-\beta\omega_0})\gamma_{\perp}(\omega_0)\Bigr) .
\label{T2_prim}
\end{equation}
\subsubsection{The case of periodic kicks at resonance and zero temperature}
\begin{figure}[htbp]
	\centering
		\includegraphics{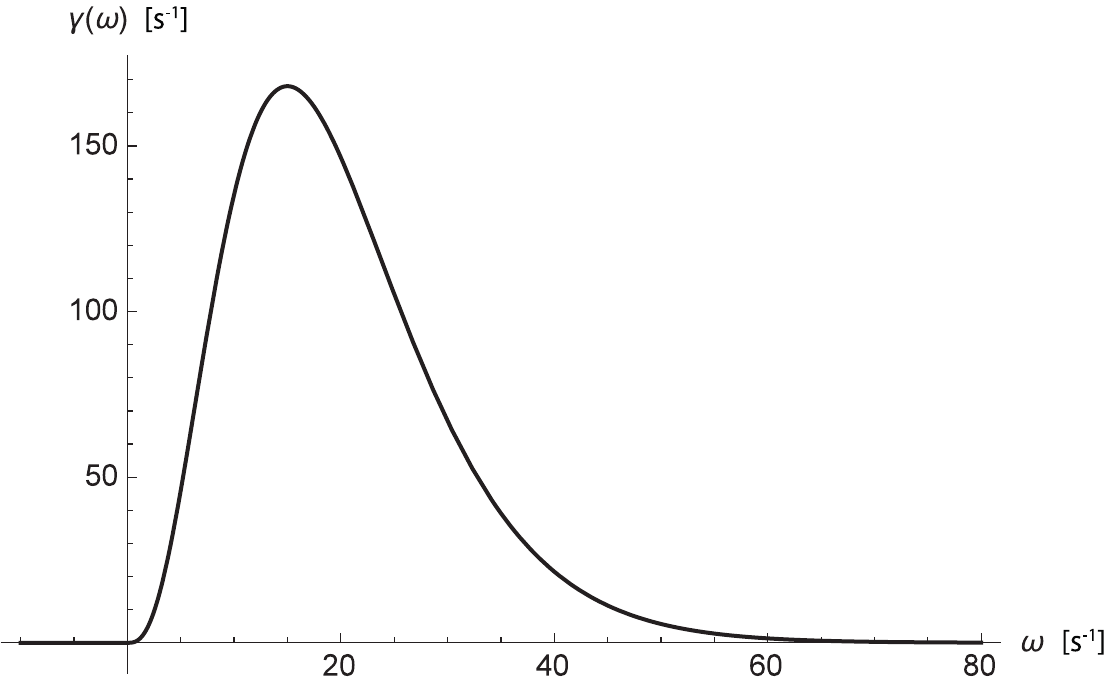}
	\caption{Example plot of spectral density function \eqref{phonon_bath} for the zero temperature case. Parameters were chosen such that $A=1$ and $\omega_{\mathrm{cut}}=5$.}
	\label{fig:SDFcutoff}
\end{figure}

Now it is possible to discuss the case of periodic kicks in a similar fashion to the previous section, using the spectral density \eqref{phonon_bath}. However, formulas obtained for the relaxation times are very complicated and involve a series of special functions. Hence, we restrict ourselves to the resonance driving and zero temperature of the bath, i.e., $\Delta = 0$, $\beta = \infty$ (Fig. \ref{fig:SDFcutoff}). This case is sufficient to present the basic feature of the periodically kicked system--suppression of dissipation and decoherence for $\Omega \gg \omega_{\mathrm{cut}}$.

\par
\begin{figure}[htbp]
	\centering
		\includegraphics{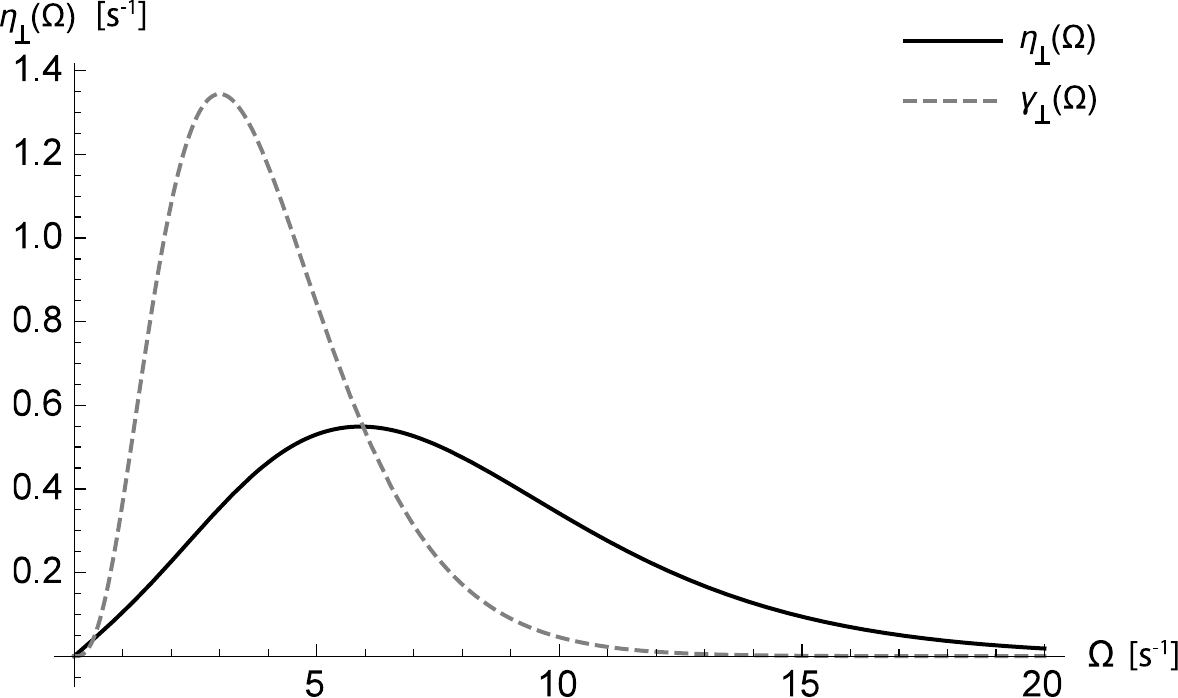}
	\caption{Case of coupling by $\pauli{1}$ and $\pauli{2}$ with $\Delta = 0$, $A = 1$ and $\omega_{\mathrm{cut}}=1$; comparison between functions $\eta_{\perp}(\Omega)$ (black) and $\gamma_{\perp}(\Omega)$ (gray, dashed).}
	\label{fig:GammaT22}
\end{figure}

The Heisenberg picture dynamics in the rotating frame of $\pauli{2}$ leads to the Fourier decomposition
\begin{equation}
		\pauli{2}(t) = \sum_{q=-\infty}^{\infty} \left( S_{+}^{2}(q) e^{it\left(\frac{\pi}{T}+q\Omega\right)} + S_{-}^{2}(q) e^{it\left(-\frac{\pi}{T}+q\Omega\right)} \right)
\end{equation}
with matrices
\begin{equation}
		\qquad S_{+}^{2}(q) = -\frac{2}{\pi} \left( \begin{array}{cc} 0 & 0 \\ \frac{1}{1+2q} & 0 \end{array} \right), \qquad S_{-}^{2}(q) = -\frac{2}{\pi} \left( \begin{array}{cc} 0 & \frac{1}{1-2q} \\ 0 & 0 \end{array} \right).
\end{equation}
given again in Floquet basis. Due to the condition $\Delta = 0$, $ \pauli{1}$ is a constant of motion and hence, because $\gamma_{\perp}(0) =0$, it does not enter the generator which, in interaction picture, reads
\begin{align}
	\mathcal{L}\rho &= \sum_{q=1}^{\infty} \gamma_{\perp}\left(-\frac{\Omega}{2}+q\Omega\right) \left(S_{-}^{2}(q)\rho S_{+}^{2}(-q) - \frac{1}{2}\acomm{S_{+}^{2}(-q)S_{-}^{2}(q)}{\rho}\right)\\
	&+ \sum_{q=0}^{\infty} \gamma_{\perp}\left(\frac{\Omega}{2}+q\Omega\right) \left(S_{+}^{2}(q)\rho S_{-}^{2}(-q) - \frac{1}{2}\acomm{S_{-}^{2}(-q)S_{+}^{2}(q)}{\rho}\right), \nonumber
\end{align}
and in the explicit matrix form is the same as \eqref{Lindblad_matrix}
\begin{equation}\label{Lindblad_matrix1}
	\lgen\rho = -\eta_{\perp}(\Omega)\left( \begin{array}{cc} \rho_{11} - \rho_{22} & -\rho_{12} \\ \rho_{21} & \rho_{22} - \rho_{11} \end{array}\right),
\end{equation}
with the single relaxation rate $\eta_{\perp}(\Omega)$ given by the formula
\begin{equation}
	\eta_{\perp}(\Omega) = \frac{A\Omega^3}{4\pi^2} \coth{\frac{\Omega}{2\omega_{\mathrm{cut}}}} \left(\sinh{\frac{\Omega}{2\omega_{\mathrm{cut}}}}\right)^{-1}.
\label{rate_perp}
\end{equation}
The plot of the relaxation rate \eqref{rate_perp} in Fig. \ref{fig:GammaT22} shows the suppression of decoherence for high kicking frequencies, $\Omega \gg \omega_{\mathrm{cut}}$.
\par
After returning to the original $\pauli{3}$ basis and to the Schr\"{o}dinger picture \eqref{eq_MME_Schroedinger}, the full dynamics of $\rho(t)$ is then expressed as [$\eta \equiv \eta _{\perp}(\Omega)$]
\begin{subequations}\label{eq:DensityOperatorSigma1Schroedinger}
	\begin{align}
		\rho_{11}(t) &= \frac{1}{2} \left[ 1+(-1)^{\floor{\frac{t}{T}}}e^{-\eta t} (\rho_{11}(0)-\rho_{22}(0)) \right], \\
		\rho_{21}(t) &= \frac{1}{2}e^{i\omega_{0}t}\Bigl[e^{-2\eta t}(\rho_{12}(0) + \rho_{21}(0)) - (-1)^{\floor{\frac{t}{T}}} e^{-\eta t} (\rho_{12}(0)-\rho_{21}(0)) \Bigr], \\
		\rho_{22}(t) &= 1 - \rho_{11}(t), \qquad \rho_{12}(t) = \overline{\rho_{21}(t)}.
	\end{align}
\end{subequations}
Equivalently, the Bloch vector representation reads
\begin{subequations}
	\begin{align}
		x_{1}(t) &= e^{-2\eta t} x_{1}(0) \cos{\omega_0 t} - (-1)^{\floor{\frac{t}{T}}}e^{-\eta t} x_{2}(0) \sin{\omega_0 t}, \\
		x_{2}(t) &= e^{-2\eta t} x_{1}(0) \sin{\omega_0 t} + (-1)^{\floor{\frac{t}{T}}}e^{-\eta t} x_{2}(0) \cos{\omega_0 t}, \\
		x_{3}(t) &= (-1)^{\floor{\frac{t}{T}}} e^{-\eta t} x_{3}(0).
	\end{align}
\end{subequations}
This evolution of Bloch vector coincides with Eq. \eqref{bloch_evolution} under the resonance condition $\Delta =0$, and again consists of three components: (a) rotation with the Larmor frequency $\omega_0$ , (b) exponential decay with two decay rates, $\eta_{\parallel}(T)$ and $2\eta_{\parallel}(T)$, and (c) inversion of the slowly decaying component performed at the times $nT$. The exponential decay of all components of Bloch vector drives the TLS to the final maximally mixed state.
\par
Assuming that the parallel and perpendicular couplings are statistically independent their combined effect is described in the interaction picture by the simple generator of the form \eqref{Lindblad_matrix} with the overall decay rate $\eta = \eta_{\parallel}+\eta _{\perp}$.

\section{Conclusions}

The formalism of Markovian master equations for periodically controlled open quantum systems weakly coupled to stationary environments allows us to revisit the well-known theory of ``bang-bang" control providing a more detailed, quantitative and mathematically consistent description. Closed formulas can be applied directly to experiments concerning periodic control for various realizations of TLS. The methods are presented in a way which can be easily extended beyond the special choice of parameters used here to illustrate the main features of the pulsed control.
\par
In the case of NMR, by comparing the measured decoherence time for two values of kicking period: $T \gg \tau_c$ and $ T \ll \tau_c$, one can determine the relaxation time of the bath, $\tau_c$, by using Eq. \eqref{decoh_rate}. This is a new parameter beside the intensity of a signal and two relaxation times: ``spin-lattice relaxation time''--$T_1$, and ``spin-spin relaxation time''--$T_2$. Such a new parameter, which depends on the environmental properties independent of those determining $T_1$ and $T_2$, could, in principle, increase the contrast in NMR imaging.

\section*{Acknowledgements}

R.A. and K.S. are supported by the Foundation for Polish Science TEAM project cofinanced by the EU European Regional Development Fund.

\appendix

\section{Notes on Floquet formalism for kicked dynamics}
\label{app_FloquetOperatorDerivation}

Here we sketch a more detailed, but not too formal derivation of the Floquet operator of a periodically kicked system as well as the evolution operator $U(t)$, justifying formulas \eqref{eq_KickedFloquetOperator} and \eqref{eq_PropagatorSawtooth}. In order to do so, we accept an approach mentioned in Ref. \cite{McCawPhDthesis:2005}, namely, we consider a Floquet operator $U(T)$ as an evolution operator, which propagates state $\psi(nT^{+})$ of a system from a time \emph{just after} the $n$th kick to a state $\psi((n+1)T^{+})$ of a system at time \emph{just after} the kick $n+1$. Here, the phrase ``just after'' is to be understood in a sense of limiting procedure, i.e., $t^{+}$ will be defined as $t + \epsilon$ for $\epsilon$ being positive and arbitrarily small. Equivalently, one can choose another time frame for propagator with time 0 chosen ``just after'' the first kick, i.e., at $t = \epsilon$. In this new time frame, all kicks are slightly displaced, by $\epsilon$, in the direction of negative $t$.
\par
It must be noted, however, that many sources use seemingly different approach \cite{McCawPhDthesis:2005,Combescure:1990,McCawEtAl:2005,Astaburuaga:2006} where $U(T)$ is calculated as the evolution from state $\psi(nT^{-})$ to $\psi((n+1)T^{-})$ with $t^{-}$ denoting time \emph{just before} time $t$, e.g., $t^{-} = t-\epsilon$ with $\epsilon$ again positive and small. This, being a rather opposite convention results in different Floquet operator, which turns out to be completely equivalent in most applications to ours.
\par
\begin{figure}[htbp]
	\centering
		\includegraphics{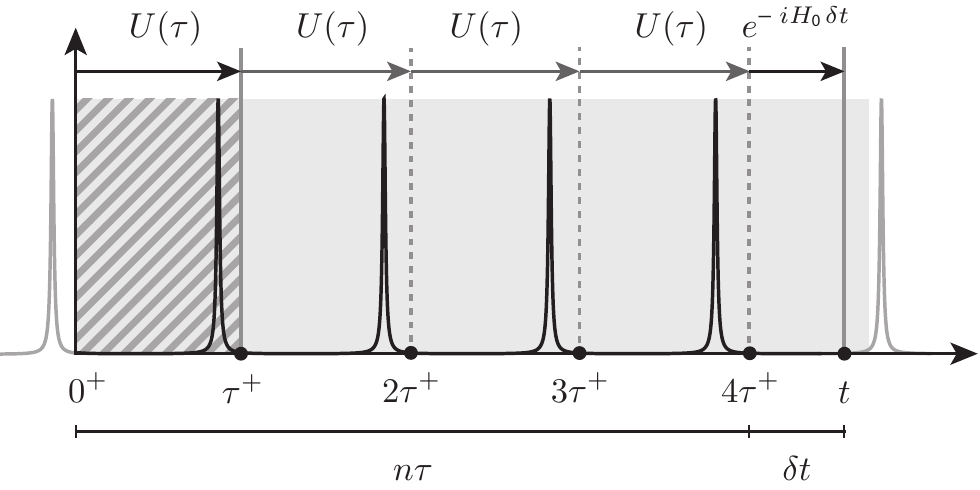}
	\caption{Schematic interpretation of evolution on the interval $[0^{+},t]$. Each kick is taking place just before time $nT$ and is included within evolution on $[nT^{+}, (n+1)T^{+}]$. A whole evolution is then expressed as a composition $U(t) = e^{-iH_{0}\delta t}U(T)^{n}$ of free evolution (generated by $H_{0}$) taking a time $\delta t$ with $n$ copies of Floquet operator, which together give the remaining evolution over time $nT$.}
	\label{fig:Figure1}
\end{figure}
\begin{figure}[htbp]
	\centering
		\includegraphics{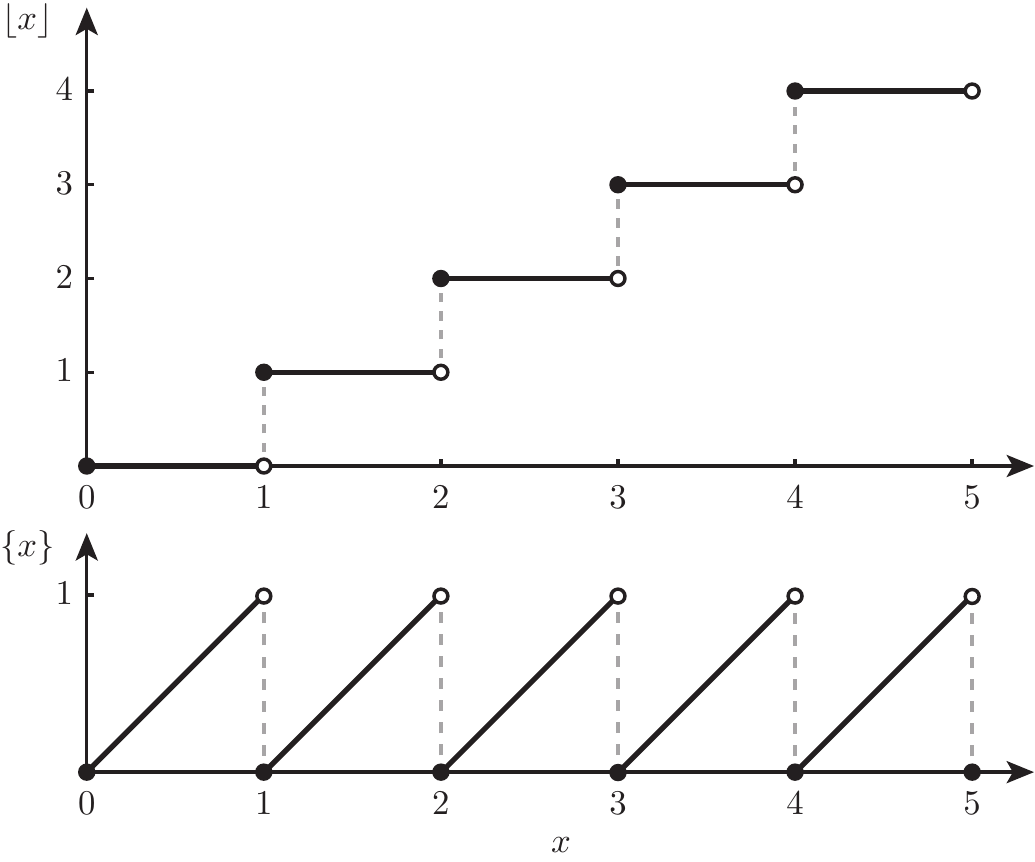}
	\caption{Plots of integer part $\floor{x}$ (up) and fractional part $\fp{x}$ (down) of $x$.}
	\label{fig:Figure2}
\end{figure}

We consider evolution operator defined over interval $[0^{+},t]$, such that the kick occuring at 0 is not counted. All subsequent kicks take place at times $nT$, $n \in \integer$. Informally speaking, each one of the $\delta$ functions in the Dirac Comb $\sum_{k=-\infty}^{\infty} \delta(t-kT)$ can be approximated by a suitable well-behaved function, nonzero over the interval $[nT^{-},nT^{+}]$, with very high and very narrow peak at $t = nT^{-}$ and being zero on the interval $[nT^{+}, (n+1)T^{-}]$ (such reasoning can be made rigorous by appropriate regularization of the $\delta$ function). It is evident, that, in $[0^{+}, T^{-}]$, the time-dependent part of Hamiltonian $H(t)$ is not acting, so the system evolves freely via $e^{-iH_{0}t}$. In the infinitesimally short interval $[T^{-},T^{+}]$, where a kick takes place, the system experiences an infinitesimally short and infinitely strong kick by operator $\lambda A$ in such a way, that evolution is rapidly modified by an ``instantaneous unitary'' of the form
\begin{equation}
	\exp{\left\{-i\int\limits_{T^{-}}^{T^{+}} \left(H_0 + \lambda A \delta(t-T) \right) \, dt\right\}} \equiv e^{-i\lambda A},
\end{equation}
where the equivalence comes from the fact, that during such short time only $\delta (t-T)$ contributes significantly and $H_0$ can be entirely omitted. Therefore, formally considering the limiting procedure one can replace $t^{+}$ and $t^{-}$ with $t$ and the evolution over interval $[0,T]$ is just a Floquet operator
\begin{equation}
	U(T) = e^{-i\lambda A} e^{-i H_{0} T}.
\end{equation}
For graphical explanation, see Fig. \ref{fig:Figure1}.
Now let us take $t$ large, i.e., $t$ can be expressed as $t = nT + \delta t$ for $n \in \natur$ and $0 \leqslant \delta t < T$. Then, one has
\begin{subequations}\label{eq_nDeltaTdefinitions}
\begin{align}
	&n = \floor{\frac{t}{T}}, \\
	&\delta t = T \left( \frac{t}{T} - \floor{\frac{t}{T}} \right) = T\fp{\frac{t}{T}}
\end{align}
\end{subequations}
with $\floor{x}$ and $\fp{x}$ denoting the integer part (floor function) and fractional part (sawtooth function) of $x$ with property $\fp{x} = x-\floor{x}$. The full evolution operator $U(t)$ can therefore be composed of a free evolution over time $\delta t$ with $n$ subsequent evolutions $U(T)$ over time intervals $[kT, (k+1)T]$, i.e.
\begin{equation}\label{eq:propagator1}
	U(t) = e^{-i H_{0} \delta t} U(T)^{n}
\end{equation}
which, after using Eq. \eqref{eq_nDeltaTdefinitions}, is
\begin{equation}
	U(t) = e^{-i H_{0} T\fp{\frac{t}{T}}} U(T)^{\floor{\frac{t}{T}}}.
\end{equation}
Of course $U(T) = e^{-i\bar{H}T}$ and $\floor{x} = x - \fp{x}$, so one has
\begin{align}\label{eq:propagator2}
	U(t) &= e^{-i H_{0} T\fp{\frac{t}{T}}} e^{-i\bar{H}T\floor{\frac{t}{T}}} \\
	&= e^{-i H_{0} T\fp{\frac{t}{T}}} e^{i\bar{H}T\left(\fp{\frac{t}{T}}-\frac{t}{T}\right)} \nonumber \\
	&= e^{-i H_{0} T\fp{\frac{t}{T}}} e^{i\bar{H}T \fp{\frac{t}{T}}} e^{-i\bar{H}t} \nonumber \\
	&= P(t) e^{-i\bar{H}t}, \nonumber
\end{align}
where operator $P(t) =e^{-i H_{0} T\fp{\frac{t}{T}}} e^{i\bar{H}T \fp{\frac{t}{T}}} $ is clearly periodic:
\begin{align}
	P(t+T) &= e^{-i H_{0} T\fp{\frac{t+T}{T}}} e^{i\bar{H}T \fp{\frac{t+T}{T}}} \\
	&= e^{-i H_{0} T\fp{\frac{t}{T}+1}} e^{i\bar{H}T \fp{\frac{t}{T}+1}} \nonumber \\
	&= e^{-i H_{0} T\fp{\frac{t}{T}}} e^{i\bar{H}T \fp{\frac{t}{T}}}, \nonumber
\end{align}
since $\fp{x+1} = \fp{x}$ (see Fig. \ref{fig:Figure2}).

\bibliography{Bibliografia}

\end{document}